\documentclass[11pt]{article}

\usepackage[utf8]{inputenc}
\usepackage{natbib}
\usepackage{amsmath}
\usepackage{amsfonts}
\usepackage{graphicx}
\usepackage{bm}
\usepackage{authblk}

\newcommand{\E}{\mathbf{E}}
\newcommand{\I}{\mathbf{1}}

\newcommand{\R}{\mathbb{R}}

\newcommand{\Btheta}{\bm{\theta}}
\newcommand{\Bbeta}{\bm{\beta}}
\newcommand{\D}{\mathop{}\!\mathrm{d}}
\newcommand{\responsepattern}{\bm{y}}
\newcommand{\covariatepattern}{\bm{x}}
\newcommand{\responseprocess}{Y}
\newcommand{\covariateprocess}{X}
\newcommand{\responsepoint}{y}
\newcommand{\covariatepoint}{x}

\graphicspath{{pictures/}}

\begin{document}

\title{Hierarchical log Gaussian Cox process for regeneration in uneven-aged forests}

\author[1]{Mikko Kuronen}
\author[2]{Aila Särkkä}
\author[3]{Matti Vihola}
\author[1]{Mari Myllymäki}

\affil[1]{Natural Resources Institute Finland (Luke)}
\affil[2]{Mathematical Sciences, Chalmers University of Technology and the University of Gothenburg}
\affil[3]{Department of Mathematics and Statistics, University of Jyväskylä}

\maketitle

\begin{abstract}

We propose a hierarchical log Gaussian Cox process (LGCP) for point patterns, where a set of points $\covariatepattern$ affects another set of points $\responsepattern$ but not vice versa. We use the model to investigate the effect of large trees to the locations of seedlings.
In the model, every point in $\covariatepattern$ has a parametric influence kernel or signal, which together form an influence field. Conditionally on the parameters, the influence field acts as a spatial covariate in the intensity of the model, and the intensity itself is a non-linear function of the parameters.
Points outside the observation window may affect the influence field inside the window.
We propose an edge correction to account for this missing data.
The parameters of the model are estimated in a Bayesian framework using Markov chain Monte Carlo (MCMC) where a Laplace approximation is used for the Gaussian field of the LGCP model.
The proposed model is used to analyze the effect of large trees on the success of regeneration in uneven-aged forest stands in Finland.

\emph{Key words: Bayesian inference, competition kernel, Laplace approximation, MCMC, spatial random effects, tree regeneration}
\end{abstract}

\section{Introduction}\label{sec:intro}

Hierarchical relationships or interactions, where a plant species affects the locations or intensity of another species but not vice versa often occur in ecological communities \citep[e.g.][]{DieckmannEtal2000}. An example of such a hierarchical relationship is that proximity of large trees affects the intensity of seedling either positively, e.g.\ by protecting against wind, or negatively by giving too much shade. Mathematically, we can describe such plant communities by two point processes, $\responseprocess$ and $\covariateprocess$, where one ($\covariateprocess$) is affecting the other ($\responseprocess$) but not vice versa.

The hierarchical interaction assumption affects the inference for $\responseprocess$ and $\covariateprocess$ greatly since $\covariateprocess$ can be modeled independently of $\responseprocess$ and $\responseprocess$ is modeled conditionally on $\covariateprocess$. A realization of the point process $\covariateprocess$ acts then as a source of heterogeneity in the distribution of $\responseprocess$. \citet{HogmanderSarkka1999} modeled interaction between two territorial ant species using Gibbs point processes under such an assumption. A similar hierarchical Gibbs point process approach was used in \citet{GrabarnikSarkka2009} and \citet{GenetEtal2014}.
Furthermore, \citet{IllianEtal2009} modeled the spatial pattern of resprouter species ($\responseprocess$) given the locations of seeders ($\covariateprocess$) in a hierarchical set-up having an inhomogeneous Poisson process as a model for the resprouters.

\begin{figure}[ht!]
\centering\includegraphics[width=\textwidth]{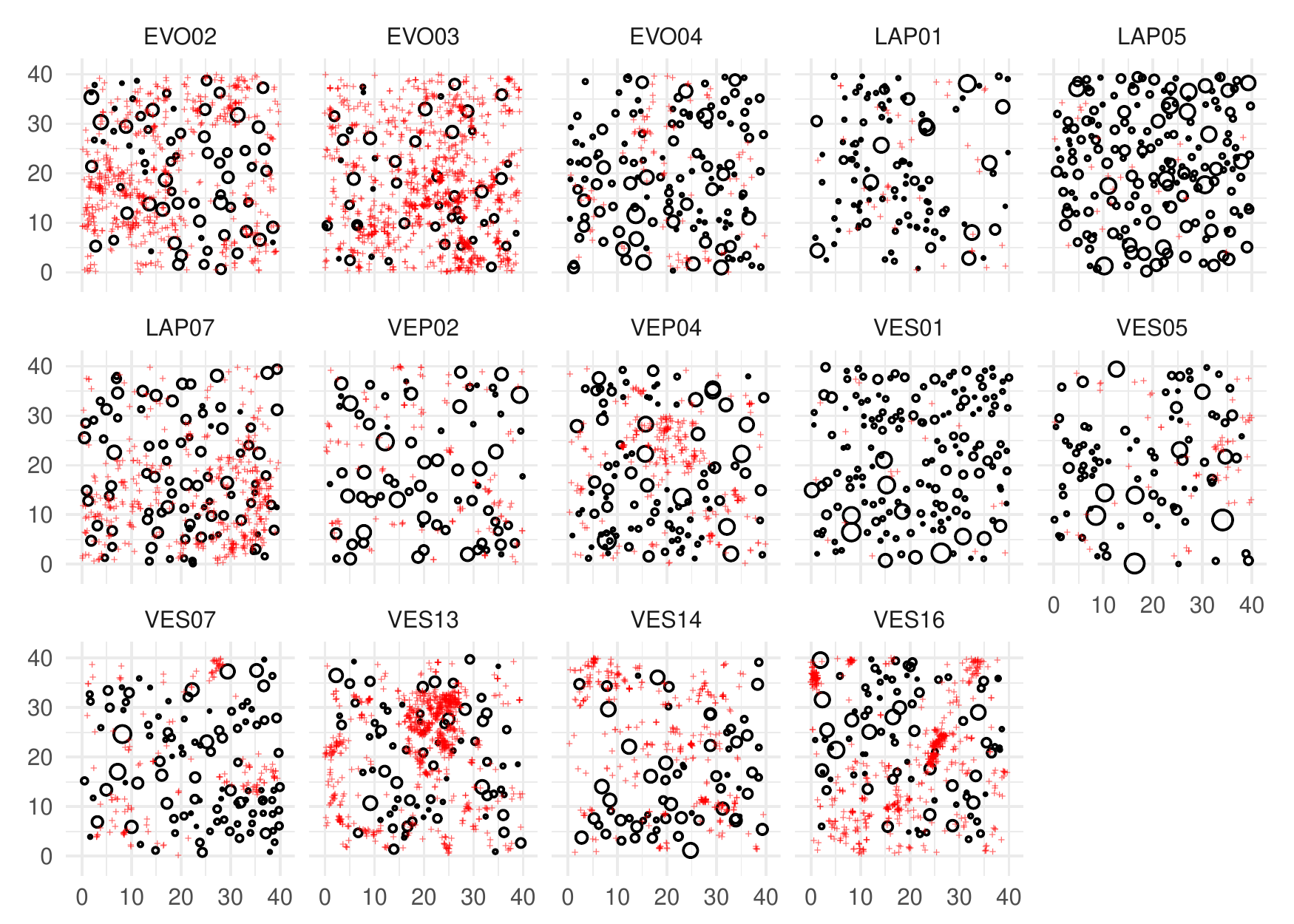}
\caption{Trees with dbh at least $7$ cm (open circles with radii relative to the dbh of the tree) and new seedlings (red crosses) in areas of size $40\text{ m}\times 40\text{ m}$. The headings give abbreviations for the plot locations and numbers.} \label{fig:seedlingppall}
\end{figure}

Here, we model the intensity of new seedlings in a spruce-dominated uneven-aged (boreal) forest given the locations and diameters at breast height (dbh) of large trees.
Thus, our $\covariateprocess$ process of large trees is a marked point process, where the mark of a tree is the dbh.
The data consist of 14 sample plots from an experiment of continuous cover forestry involving single-tree selection in four nearby areas in Southern Finland (Figure \ref{fig:seedlingppall}).
The system relies on natural emergence of new seedlings and
a continuous recruitment is necessary for long-term sustainability in a wide sense \citep[e.g.][]{EerikainenEtal2014, KuusinenEtal2019}.
While a sufficient number of seedlings is necessary for the success of regeneration, our focus here is in the spatial distribution of the seedlings within the plots, and the effect of large trees on it.

Like in the resprouter and seeder case above, an inhomogeneous Poisson process would be a reasonable model since the effect of large trees could be added in the model as an explanatory variable.
However, already visual inspection of the patterns of seedlings $\responsepattern$ indicates that the patterns tend to be rather clustered, beyond the clustering that may be explained by the patterns of large trees $\covariatepattern$. Due to such
unexplained clustering, a log Gaussian Cox process (LGCP) \citep{MollerEtal1998} is a more appropriate model for the conditional point process of seedlings given large trees.

To model the effect of the large trees $\covariateprocess$, we assume that each tree $\covariatepoint\in \covariateprocess$ emits a signal or impulse
that describes the effect of the tree to its neighborhood.
We assume that this effect decreases with the distance from the tree $\covariatepoint$.
In general, the size of the effect as well as the range of the effect could depend on the size or other properties of the tree, e.g.\ its dbh.
Because we do not have precise a priori information on the size and range of the effects,
we use parametric signals similar to the ones found in the literature \citep{Adler1996, PommereningEtal2011b, HabelEtal2019b, PommereningGrabarnik2019}.
The individual signals are then superimposed to form an influence field, which describes the overall influence of the points of $\covariateprocess$ on any location $s$ in the observation window $W$.
These kind of models have been used to model, for example, effect of neighboring individuals on the growth of a subject tree, survival of seedlings and ground vegetation in different contexts \citep[e.g.][]{WuEtal1985, MiinaPukkala2002, PommereningEtal2011b, HabelEtal2019b, KuuluvainenPukkala1989, KuhlmannBerenzonEtal2005}.

Our idea here is to include the superimposed individual signals in the log intensity function of the LGCP model.
Using parametric models for the signals, the intensity of the LGCP is a non-linear function of the model parameters.
According to \cite{PommereningSanchezMeador2018} the signals are aggregated additively or multiplicatively and there is no evidence to prefer either of these ways.
We follow \cite{PommereningEtal2011b} and \citet{IllianEtal2009} and aggregate the signals additively.

Our Bayesian inference algorithm is based on Markov chain Monte Carlo (MCMC) sampling for parameters, and a Laplace approximation is used for the latent random field of the LGCP to avoid high-dimensional MCMC sampling.
Laplace approximations are widely used for inference of latent Gaussian fields, for instance within the popular INLA method \citep{RueEtal2009}.
However, in contrast to INLA, the MCMC is more robust, and can cope with multimodal parameter posteriors.

The large tree process typically extends beyond the borders of the sample plot. However, we have observed the process in the same observation window as the seedlings.
Thus, the influence field computed only from the observed trees is weaker near the borders than the field computed from the fully observed large tree process would be. In order to account for the unobserved trees outside the observation window,
we compute the influence field using an edge correction method similar to that suggested in \citet{KuhlmannBerenzonEtal2005}: the unobserved trees are imputed based on the assumption that the locations of large trees are distributed according to a Poisson process.
This rather simple edge correction method can be efficiently implemented within the Bayesian inference, in contrast to alternatives where the locations (and sizes) of unobserved large trees would be included in the Bayesian inference as unknowns and simulated within the MCMC approach.

The rest of the paper is organized as follows. In Section \ref{sec:model}, we give some examples of influence kernels and introduce the conditional LGCP model.
The Bayesian estimation approach including the edge correction is described in Section \ref{sec:inference}. Section \ref{sec:simstudy} presents the results of a simulation experiment that was conducted to explore the performance of the proposed estimation and edge correction methods.
Finally, the forestry data are described in further detail and studied in Section \ref{sec:application}. Section \ref{sec:discussion} is for discussion.

\section{Conditional log Gaussian Cox process model}\label{sec:model}

Let us have a bivariate point process in $\mathbb{R}^2$ consisting of an unmarked point process $\responseprocess$ and an unmarked or a marked point process $\covariateprocess$.
Let us further assume that we have observed a realization of process $\responseprocess$, namely $\responsepattern=\{\responsepoint_i\}$, in a bounded window $W\subset \mathbb{R}^2$.
Our primary interest is in the spatial pattern $\responsepattern$ which is affected by
a realization $\covariatepattern$ of the spatial point process $\covariateprocess$.
The spatial pattern $\covariatepattern$ can consist only of the point locations $x_j$ or of the point locations and marks, $[x_j,m_j]$, if some characteristics (marks) $m_j$ of the points $x_j$ are available.
In our forestry application, $\responsepattern$ consists of the locations of seedlings, while $\covariatepattern$ is the pattern of locations and dbh's of large trees.

In our approach, the effect of $\covariatepattern$ on $\responsepattern$ is modeled using the influence kernels around the points of $\covariatepattern$ that are explained in Section \ref{sec:competition field}.
To account for the clustering in the pattern $\responsepattern$ not explained by $\covariatepattern$,
the LGCP model is proposed and defined in Section \ref{sec:condmodel}.
Replicated point patterns are discussed in Section \ref{sec:replicates}.

\subsection{Influence kernels and influence field}\label{sec:competition field}

We assume that each point $[\covariatepoint_j,m_j]$ of the process $\covariateprocess$ introduces an influence kernel around its location.
We focus on isotropic influence kernels of the form
$c(h; m_j, \Btheta_I)$,
where $h=\|s-\covariatepoint_j\|$ is the distance between the location $s$ of interest and $\covariatepoint_j$.
Many kernels have been suggested in the literature for different applications \citep[e.g.][]{Adler1996, IllianEtal2008, PommereningEtal2011b, PommereningMaleki2014, SchneiderEtal2006}.
We used a mark independent Gaussian kernel
\begin{equation}\label{kernel_gaussian}
c(h; \theta) = \exp\left(-(h/\theta)^2\right),
\end{equation}
where $\theta>0$ is an unknown influence range parameter.
Here the influence of a point gradually decreases with the distance from the point.

A mark dependent generalization of (\ref{kernel_gaussian}) is given by
\begin{equation}\label{kernel_Arne}
c(h, m; \Btheta_I) = m^{\alpha} \exp\left( - \left(\frac{h}{\theta m^{\delta}} \right)^2 \right)
\end{equation}
with $\Btheta_I=(\theta, \delta, \alpha)$, where $\theta > 0$, $\delta > 0$, and $\alpha \ge 0$.
If $\alpha = 0$, the mark affects only the range of influence and if $\alpha > 0$, it affects both the range and the strength \citep[e.g.][]{PommereningEtal2011b}.

The influence field of the process $\covariateprocess$ can then be defined as a superposition of the individual influence kernels,
\begin{equation*}
C(s; \Btheta_I, \covariateprocess) = \sum_{[\covariatepoint_j,m_j] \in \covariateprocess} c(\|s-\covariatepoint_j\|, m_j; \Btheta_I).
\end{equation*}

\subsection{Conditional model}\label{sec:condmodel}

Since $\responsepattern$ is affected by $\covariatepattern$, we introduce a conditional point process model for $\responsepattern$ given $\covariateprocess=\covariatepattern$, where the intensity of $\responseprocess$ is affected by the influence field of $\covariatepattern$.
This conditional model is a LGCP with the intensity
\begin{equation}\label{LGCP_Lambda}
\Lambda(s; \Bbeta, \Btheta_I, \covariatepattern, Z) = \exp( \beta_0 + \beta_1 C(s; \Btheta_I, \covariatepattern) + Z(s) ),
\end{equation}
where $C(s; \Btheta_I, \covariatepattern)$ is a parametric influence field, $\Bbeta=(\beta_{0}, \beta_1)$ and the unknown coefficients $\beta_0\in \R$ and $\beta_1\in \R$ are the intercept and the strength of the influence field, respectively. If $\beta_1<0$, $\covariatepattern$ affects the intensity of $\responseprocess$ negatively and the influence field $C(s;\Btheta_I, \covariatepattern)$ can be interpreted as a thinning of the LGCP process with intensity $\Lambda(s) = \exp(\beta_0 + Z(s))$.
If, however, $\beta_1>0$, $\covariatepattern$ has a positive effect on the intensity of $\responseprocess$ and there are more points of  $\responseprocess$ in areas with a high value of $C(s;\Btheta_I, \covariatepattern)$. Furthermore,
$Z := \{Z(s): s\in\mathbb{R}^2\}$ is a zero-mean stationary Gaussian random field with a covariance function $C_Z(r; \Btheta_Z)$ and independent of the influence field.
In our application below, we use the Mat{\'e}rn covariance function
\begin{equation}\label{MaternCf}
C_Z(r; \Btheta_Z, \nu)
= \sigma_Z^{2} \frac {2^{1-\nu}}{\Gamma(\nu)} \left(\sqrt{2\nu}\frac{r}{\rho_Z}\right)^{\nu}K_{\nu} \left(\sqrt{2\nu}\frac{r}{\rho_Z}\right), \quad r>0,
\end{equation}
with the smoothness parameter $\nu=2$ and $\Btheta_Z = (\sigma_Z^2, \rho_Z)$, where
$\sigma_Z^2$ and $\rho_Z$ are the variance and range parameters, respectively, and $K_\nu$ is the modified Bessel function of the second kind \citep[e.g.][]{Cressie1993, ChilesDelfiner1999, BanerjeeEtal2004}.
The choice $\nu=2$ was made since we
expect that the unobserved environmental conditions that affect the clustering of $\responsepattern$ in our application vary rather smoothly
and since it is computationally convenient \citep{LindgrenEtal2011}.

\subsection{Replicates}\label{sec:replicates}

Assume that we have several independent replicated point patterns $\responsepattern_k$, $k = 1, \dots, N$, from the conditional distribution of the point process $\responseprocess$ given $\covariateprocess=\covariatepattern_k$, $k = 1, \dots, N$.
Conditionally on $\covariateprocess=\covariatepattern_k$,
the model for $\responsepattern_k$ is a LGCP with the intensity
$\Lambda(s; \beta_{0}, \beta_{1}, \Btheta_I, \covariatepattern_k, Z_k)$ in (\ref{LGCP_Lambda}),
where $Z_k$, $k = 1, \dots, N$, are independent replicates of the Gaussian random field with parameters $\Btheta_Z$.
For our data, it is not reasonable to assume that all replicates have the same $\beta_{0}$, which controls the number of points of $\responseprocess$, and we let each pattern $\responsepattern_k$ have its own intercept parameter $\beta_{0}$, i.e.\ $\beta_{0k}$ for $\responsepattern_k$, $k=1,\dots,N$.
Consequently, in our application below, the pattern $\responsepattern_k$ is assumed to be a realization of the LGCP model with the intensity $\Lambda(s; \beta_{0k}, \beta_{1}, \Btheta_I, \covariatepattern_k, Z_k)$.

\section{Inference}\label{sec:inference}

The likelihood of the conditional LGCP model for a point pattern $\responsepattern$ with $n$ points observed in $W$ is
\begin{equation}\label{eq:likelihood1}
p(\responsepattern; \Bbeta, \Btheta_I, \Btheta_Z, \covariatepattern) =
\E_{\Btheta_Z} \prod_{i = 1}^{n} \Lambda(\responsepoint_{i}; \Bbeta, \Btheta_I, \covariatepattern, Z) \exp\left(-\int_{W} \Lambda(u; \Bbeta, \Btheta_I, \covariatepattern, Z){\rm d}u\right),
\end{equation}
where $\Bbeta$, $\Btheta_I$, $\Btheta_Z$ are the model parameters, $Z$ denotes the Gaussian random field and the expectation is over $Z$ given $\Btheta_Z$.
As we use Bayesian inference we need to be able to evaluate the likelihood \eqref{eq:likelihood1} efficiently.
Below, we describe the approximations needed: discretization of the observation window (Section \ref{sec:inf_discre}), an edge-corrected influence field (Section \ref{sec:edge}), and approximations related to the Gaussian field (Section \ref{sec:inf_Z}), which include approximating the field by a Gaussian Markov random field and using the Laplace approximation to evaluate the likelihood. Finally, the approximated likelihood based on replicates is given in Section \ref{sec:inf_replicates} and the MCMC algorithm is described in Section \ref{sec:implementation}.

\subsection{Discretization}\label{sec:inf_discre}

To be able to make inference on LGCP models, the observation window $W$ of the point pattern $\responsepattern$ is discretized using a regular grid in a similar manner as in \citet{RueEtal2009} and \citet{MollerEtal1998}.
Namely, the observation window $W$ is divided into $G$ disjoint cells $\{w_g\}$ with center locations $\xi_g$ and area $A$.
Furthermore, we let $n^{\responsepoint}_g$ denote the number of observations $\responsepattern$ within $w_g$ in $W$ and $\mathbf{n}^{\responsepoint} = (n^{\responsepoint}_1,\dots,n^{\responsepoint}_G)$.
A piecewise constant approximation is used for the Gaussian field $Z$ and the competition field $C$ and the locations of $\responsepattern$ are replaced by the counts $n^{\responsepoint}_g$.
The approximate likelihood for $\mathbf{n}^{\responsepoint}$ is
\begin{equation}\label{eq:lik}
p(\mathbf{n}^{\responsepoint}; \Bbeta, \Btheta_I, \Btheta_Z, \covariatepattern) = \E_{\Btheta_Z} p(\mathbf{n}^\responsepoint; \Bbeta, \Btheta_I, \covariatepattern, Z^D),
\end{equation}
where
\begin{equation*}
p(\mathbf{n}^{\responsepoint}; \Bbeta, \Btheta_I, \covariatepattern, Z^D) = \prod_{g=1}^G \text{Pois}(n^{\responsepoint}_g; \Lambda_g),
\end{equation*}
$\Lambda_g = A\exp( \beta_0 + \beta_1 C^D(\xi_g; \Btheta_I, \covariatepattern) + Z^D(\xi_g; \Btheta_Z) )$, and $C^D$ and $Z^D$ are the piecewise constant approximations of $C$ and $Z$.

\subsection{Edge correction}\label{sec:edge}

The large tree process $X$ is only partially observed and generating the influence field only based on the observed large trees would result in too weak influence near the borders.
Therefore, we propose an imputation type approach, similar to the one proposed by \cite{KuhlmannBerenzonEtal2005}, to correct for the unobserved points of $\covariateprocess$.
Specifically we propose to replace the influence generated by the unobserved trees with the expected influence generated assuming that the whole process $\covariateprocess$ is an independently marked homogeneous Poisson process. In the unmarked case, $\covariateprocess$ is assumed to be a homogeneous Poisson process. In general the point pattern outside the window would depend on the pattern inside the window, but this is not the case for the Poisson process.

Let $\lambda$ and $F$ be the intensity and mark distribution of $\covariateprocess$, and $X_{W^c}$ the restriction of $X$ to $W^c$, the complement of $W$.
Using the Campbell theorem \citep[e.g.][]{ChiuSKM2013} we can write
\begin{align*}
    \E C(s; \Btheta_I, \covariateprocess_{W^c})
    &= \E \sum_{[\covariatepoint_j,m_j] \in \covariateprocess_{W^c}} c(\|s-\covariatepoint_j\|, m_j; \Btheta_I) \\
    &=\int_{R_+}\int_{R^2} c(s-x, m; \Btheta_I) \I_{W^c}(x) \lambda \D x \D F(m),
\end{align*}
where $\I_{W^c}$ is the indicator function of the set $W^c$, i.e.\ $\I_{W^c}(x)=1$ if $x\in W^c$, and 0 otherwise.
By changing the order of the integrals we find that
\begin{align*}
    \E C(s; \Btheta_I, \covariateprocess_{W^c})
    &= \int_{R^2} f(s-y) \I_{W^c}(y) \lambda \D y\\
    &= \int_{R^2} f(s-y) \lambda \D x - \int_{R^2} f(s-x) \I_W(x) \lambda \D x,
\end{align*}
where $f(x) = \int_{R_+}c(\|x\|, m; \Btheta_I)\D F(m)$. By changing to polar coordinates and with a slight abuse of notation
\begin{equation*}
    \int_{R^2} f(s-x) \lambda \D x = \lambda 2\pi\int_0^\infty rf(r) \D r,
\end{equation*}
which can be computed using numerical integration.
Since we are only interested in locations $s \in W$, we can replace the function $f$ with $f\I_{W^S}$, the restriction of $f$ to the set $W^S = \{s-x: s \in W, x \in W\}$, and
\begin{equation*}
    \int_{R^2} f(s-x) \I_W(x)\D x
    = \int_{R^2} (f\I_{W^S})(s-x) \I_W(x) \lambda \D x = (f\I_{W^S} * \I_W)(s).
\end{equation*}
The discrete convolution of the piecewise constant approximations of $f\I_{W^S}$, and $\I_W$
can be efficiently computed using discrete Fourier transforms \citep{OppenheimEtal1999, FFTW05}.
For $F$, we use the empirical distribution of marks in the sample plot under study.

The edge-corrected influence field value at any location $s\in W$ is then obtained as the sum of the influence field calculated from the observed $\covariatepattern_{W}$, $C(s; \Btheta_I, \covariatepattern_{W})$, and the expected influence load of the unobserved $\covariateprocess_{W^c}$.
In general, we use the numerical approximation explained above but for the special case of
the Gaussian influence kernel \eqref{kernel_gaussian} and a rectangular observation window, it is easy to compute the edge correction by hand.

\subsection{Approximations related to the Gaussian field}\label{sec:inf_Z}

We use Laplace approximation \citep{TierneyKadane1986,RueEtal2009} to approximate the likelihood \eqref{eq:lik} and obtain
\begin{equation}\label{eq:approximatelikelihood}
    \E_{\Btheta_Z} p(\mathbf{n}^\responsepoint; \Bbeta, \Btheta_I, \covariatepattern, Z^D)
    \approx \sqrt{\frac{(2\pi)^d}{\det(-\mathbf{H}(\hat{\mathbf{z}}))}} p(\mathbf{n}^\responsepoint; \Bbeta, \Btheta_I, \covariatepattern, \hat{\mathbf{z}})p(\hat{\mathbf{z}} ; \Btheta_Z),
\end{equation}
where $\mathbf{H}$ and $\hat{\mathbf{z}}$ are the Hessian and maximizer of $\log p(\mathbf{n}^\responsepoint; \Bbeta, \Btheta_I, \covariatepattern, \mathbf{z})p(\mathbf{z} ; \Btheta_Z)$, respectively, and $p(\mathbf{z} ; \Btheta_Z)$ is the probability density of the vector $\mathbf{Z}^D$ which contains the values of $Z^D$ at grid cells.

Since the Gaussian random field $Z$ is assumed to have mean zero and the Mat{\'e}rn covariance function \eqref{MaternCf} with $\nu = 2$, we can
utilize the explicit link between Gaussian fields and Markov random fields
\citep{LindgrenEtal2011}, which tells us that the distribution of $\mathbf{Z}^D$ should be approximated with a Gaussian distribution with a precision matrix given by \citet{LindgrenEtal2011}.

\subsection{Replicates}
\label{sec:inf_replicates}
Since the point patterns are assumed to be conditionally independent, the likelihoods \eqref{eq:likelihood1} for each replicate $\responsepattern_k$ can be multiplied to yield the final likelihood
\begin{equation}\label{eq:likelihood2}
p(\responsepattern_1, \dots, \responsepattern_N ;\Bbeta, \Btheta_I, \Btheta_Z, \covariatepattern_1, \dots, \covariatepattern_N)
= \prod_{k=1}^Np(\responsepattern_k; \beta_{0k}, \beta_1, \Btheta_I, \covariatepattern_k),
\end{equation}
where now $\Bbeta$ contains all the regression coefficients, i.e.\ $\Bbeta = (\beta_{01}, \dots, \beta_{0N}, \beta_1)$.
To obtain an approximation of \eqref{eq:likelihood2}, the approximations \eqref{eq:lik} and \eqref{eq:approximatelikelihood} are applied to each pattern separately.

\subsection{MCMC}\label{sec:implementation}

Combining the likelihood \eqref{eq:likelihood2} with the prior $p(\Bbeta, \Btheta_I, \Btheta_Z)$ yields the approximate posterior distribution.
To sample from the this distribution, we use Robust Adaptive Metropolis algorithm \citep{Vihola2012,Vihola},
which uses a Gaussian random-walk proposal distribution, whose covariance is updated adaptively. The limiting proposal covariance matches the shape of the posterior, such that an average acceptance rate of 0.234 is attained, following the theoretical findings
presented e.g.\ in \citet{RobertsGelmanGilks1997}.

\section{Simulation experiment}\label{sec:simstudy}

We made a simulation experiment to study the performance of the inference approach and the edge correction method suggested above. The point pattern $\covariatepattern$ was a realization of either a Poisson process or a regular Strauss process. The Strauss process \citep[e.g.][]{IllianEtal2008} was included to see whether the edge correction based on the Poisson assumption of $\covariateprocess$ would work even in a more regular case. We did not include any cluster process since in our application, the large tree patterns $\covariatepattern$ are regular. Also, based on a small simulation study (results not shown here), it is unlikely that the Poisson correction would work well when the $\covariatepattern$ pattern is strongly clustered. We did not include marks in the simulation experiment.

\subsection{Set-up}\label{sec:simstudy_setup}
The intensity parameters of the Poisson and Strauss processes were chosen such that they result in approximately 60 points in the observation window $W=[0,40]\times[0,40]$.
In the Strauss process \citep[parametrized as in][]{spatstat2015}, the intensity related parameter was 0.06, the interaction strength 0.1, and the interaction radius 2, making the resulting patterns rather regular. The $\responsepattern$ patterns were generated on $W$ and the $\covariatepattern$ patterns on the extended window $W_{\text{ext}}=[-20,60]\times[-20,60]$ to be able to use plus sampling which represents the ideal situation where no imputation is needed as the complete pattern is known.
The Gaussian kernel \eqref{kernel_gaussian} was used as the influence kernel.
Initially, the parameters of the competition field and of the Gaussian field were set to the estimates found in Section \ref{sec:application} and the intercept $\beta_0$ was chosen such that the resulting LGCP model would have 600 points on average.
First we used the estimated values $\beta_1=-0.7$ and $\theta=2.1$, called "estimated" in Figure \ref{fig:simstudy_all}.
In addition, we used either the values $\beta_1=-3$, and $\theta=2.1$ corresponding to a much stronger effect of the influence kernel ($\beta_1$) ("strong" in Figure \ref{fig:simstudy_all}) or the values $\beta_1=-0.7$, and $\theta=6$ corresponding to a much larger range of influence $\theta$ ("wide" in Figure \ref{fig:simstudy_all}) than in the data.
In all cases, $\sigma_Z=1.6$ and $\rho_Z=2.6$.
We generated 100 replicates of each $\covariateprocess$ process and one $\responsepattern$ pattern for each $\covariatepattern$.
The random intensity of the Cox process was approximated by a piecewise constant function using 0.1 m $\times$ 0.1 m cells.

We fitted the conditional LGCP model to the simulated point patterns. We discretized the observation windows to pixels of size 1 m $\times$ 1 m and set weakly informative independent priors for all model parameters as follows:
For the parameters in $\Bbeta$, we used Gaussian distributions with mean zero and standard deviation 10. For the range parameters $\rho_Z$ and $\theta$, very small and very large values do not make sense based on the discretization and window used. Thus, we set the prior to be the Gamma distribution with shape parameter 2.4 and scale parameter 1.8, implying that approximately 90\% of the prior probability is between 1 m and 10 m. Furthermore, the prior for the standard deviation of the Gaussian field $\sigma_Z$ was the exponential distribution with expectation 10, slightly favoring small values.

For each point pattern, we then ran the MCMC scheme using a) no edge correction, b) the Poisson edge correction and c) plus sampling edge correction with 100\,000 updates using the true parameter values as the starting values. For each chain we discarded 20\,000 first samples as burn-in and saved every 10th sample.
When influence was strong, most chains converged and mixed well.
However, there were problems with mixing if the influence was not so strong. In this case the effective sample size was estimated to be less than 1000 in half of the chains.
Upon closer inspection multi-modality was often the cause.
We used posterior means of each chain in the comparisons. Using posterior modes led to identical conclusions.

\subsection{Results}

\begin{figure}
\centering\includegraphics[width=\textwidth]{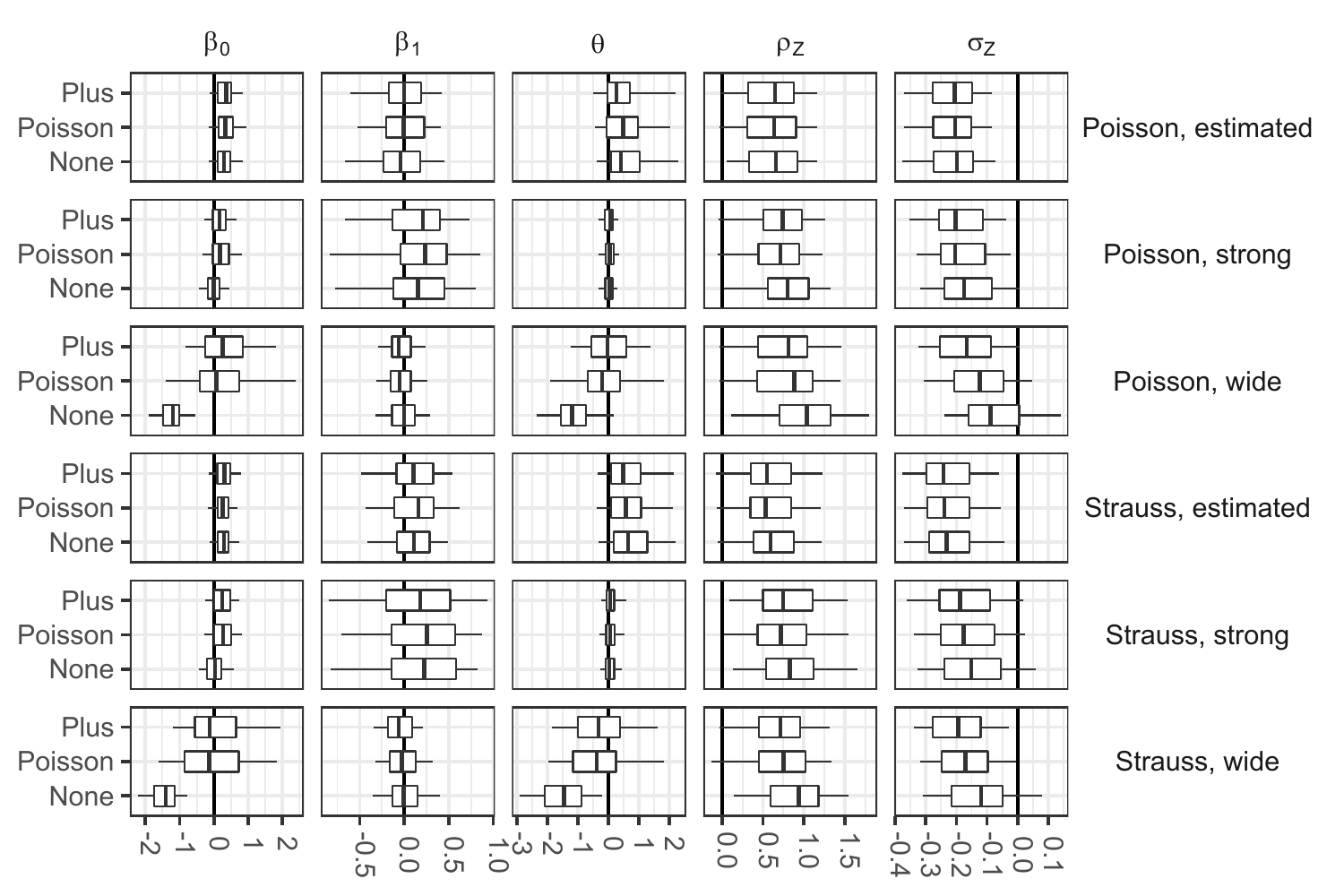}
\caption{%
Quantiles  (0.05,  0.25,  0.5,  0.75,  0.95)  of  differences  between  posterior means and reference values.
For each row of the figure, we display the $\covariateprocess$ process and the competition effect on the right and within each subfigure, we label the three different edge corrections (left). The quantiles are based on 100 replicates.
}
\label{fig:simstudy_all}
\end{figure}

First, we investigated the performance of the Bayesian inference approach. To avoid edge effects, we estimated the parameters using plus sampling, utilizing the true pattern $\covariatepattern$ in the extended window. Based on the distributions of the posterior means for the plus sampling method (see Plus in Figure \ref{fig:simstudy_all}), we can see that the Bayesian MCMC approach with the approximations used performed reasonably
well for the main parameters $\beta_1$ and $\theta$.
However, the less interesting random field parameters were clearly biased.
As expected, the distribution of the $\covariateprocess$ pattern did not affect the performance of the inference.

\begin{figure}
\centering\includegraphics[width=0.9\textwidth]{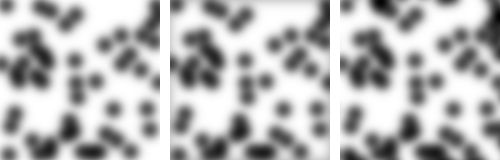}
\centering\includegraphics[width=0.9\textwidth]{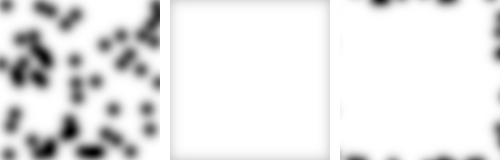}
\caption{Top row: Expected intensity of the conditional LGCP with parameters estimated from the EVO02 pattern using no edge correction (left), Poisson correction (middle) and plus sampling (right).
Bottom row: Influence field induced by the observed points (left), expected influence field caused by the unobserved points under the Poisson assumption (middle) and influence field caused by the unobserved points (right).
The $\covariatepattern$ pattern is a realization of a Strauss process with interaction parameter 0.1, interaction range 2, and with on average 60 points. Dark color means low intensity/high influence.}
\label{fig:edge_Strauss}
\end{figure}

Second, we investigated the performance of the Poisson edge correction. An example of the expected intensity field with and without edge correction for the conditional LGCP model with the parameters estimated from the EVO02 pattern and Strauss pattern $\covariatepattern$ is shown in top row of Figure \ref{fig:edge_Strauss}.
It can be seen that the Poisson corrected and the plus sampling corrected intensities are quite similar to each other.
The bottom row of Figure \ref{fig:edge_Strauss}
further shows the components of the influence field for the Poisson correction, namely the contribution of the observed points (left) and the expected contribution of the unobserved points under the Poisson assumption (middle).
The contribution of unobserved points is shown for comparison (right).
The Poisson correction simply approximates the contribution of the unobserved points.

To assess the performance of the proposed edge correction method, we compared the posterior means of the model parameters $\beta_0, \beta_1 \text{ and } \theta$, obtained by using plus sampling to the estimates obtained by using the Poisson correction and those obtained by using no edge correction. The distribution of the posterior means is shown in
Figure \ref{fig:simstudy_all}.
It can be seen that the estimates of the different methods are very similar when the influence of the large trees was not too wide, for both $\covariateprocess$ processes. However, when the influence was wide, the proposed Poisson correction produced estimates that were closer to the plus sampling based estimates than the uncorrected estimates were.
The results were altogether very similar for the Poisson and Strauss processes.
Thus, the edge correction plays a role if the range of influence of the $\covariatepattern$ points on the intensity of $\responseprocess$ is wide.

\section{Application}\label{sec:application}

The data shown in Figure \ref{fig:seedlingppall} have been collected on 40 m $\times$ 40 m squares in
southern Finland. They are part of a larger data set collected for studies on tree and stand development in managed, uneven-aged Norway spruce forests conducted under the ERIKA research project at the Natural Resources Institute Finland \citep{EerikainenEtal2007, EerikainenEtal2014, SaksaValkonen2011}.
Using the conditional LGCP model, we studied the effect of large trees $\covariatepattern_i$ (black circles) to the seedling patterns $\responsepattern_i$ (red crosses). The patterns $\covariatepattern_i$ consist of trees which had a vital crown with no damages and with a dbh at least 7 cm in 1991. Most trees (78\% of trees, 70\% basal area) were Norway spruces and the remaining ones either Scots pines or broadleaves.
The seedlings were naturally generated  with height at least $10$ cm in 1996
and had reached this height after the data collection in 1991. The seedlings were mostly Norway spruces (98\%).

We fitted the conditional LGCP model using different mark dependent and mark independent Gaussian influence fields: the full mark dependent model \eqref{kernel_Arne}, the two reduced models where either of the mark specific parameters, namely $\delta$ or $\alpha$ were set to zero, the mark independent model \eqref{kernel_gaussian}, and a model without an influence field.
The mark was always the dbh.
We used the same discretization of the observation window (1 m $\times$ 1 m pixel size) and the same priors as in the simulation experiment (Section \ref{sec:simstudy_setup}).
The pixel size 1 m $\times$ 1 m was chosen because variations in smaller units are practically unimportant in forests.
The priors for $\alpha$ and $\delta$ were both the exponential distribution with expectation 10.
We then ran the MCMC scheme using the Poisson edge correction with 120\,000 updates, leaving out the first 20\,000 observations of the chains as the burn-in.

To compare the models, we used the posterior predictive model assessment based on various summary characteristics, namely the $L$-function (variance stabilizing version of Ripley's $K$), the empty space function $F$, and the nearest neighbour distribution function $G$ summarizing the spatial pattern $\responsepattern$ and, to investigate the relationship between the large trees and seedlings, the cross $L$-function, $L_{12}$ \citep[e.g.][]{IllianEtal2008, Diggle2013}. We used the standard estimators of these functions with translational ($L$, $L_{12}$) and Kaplan-Meier edge correction methods ($F$, $G$) \citep{BaddeleyGill1997}.
For each plot, we generated 10\,000 patterns of seedlings from the posterior predictive distributions of the conditional LGCP models given the observed $\covariatepattern$ and calculated the summary functions for the data and for each of the generated patterns.
The posterior predictive simulations were made using a discretization with 0.2 m $\times$ 0.2 m cell size.

\begin{figure}[ht!]
\centering\includegraphics[width=\textwidth]{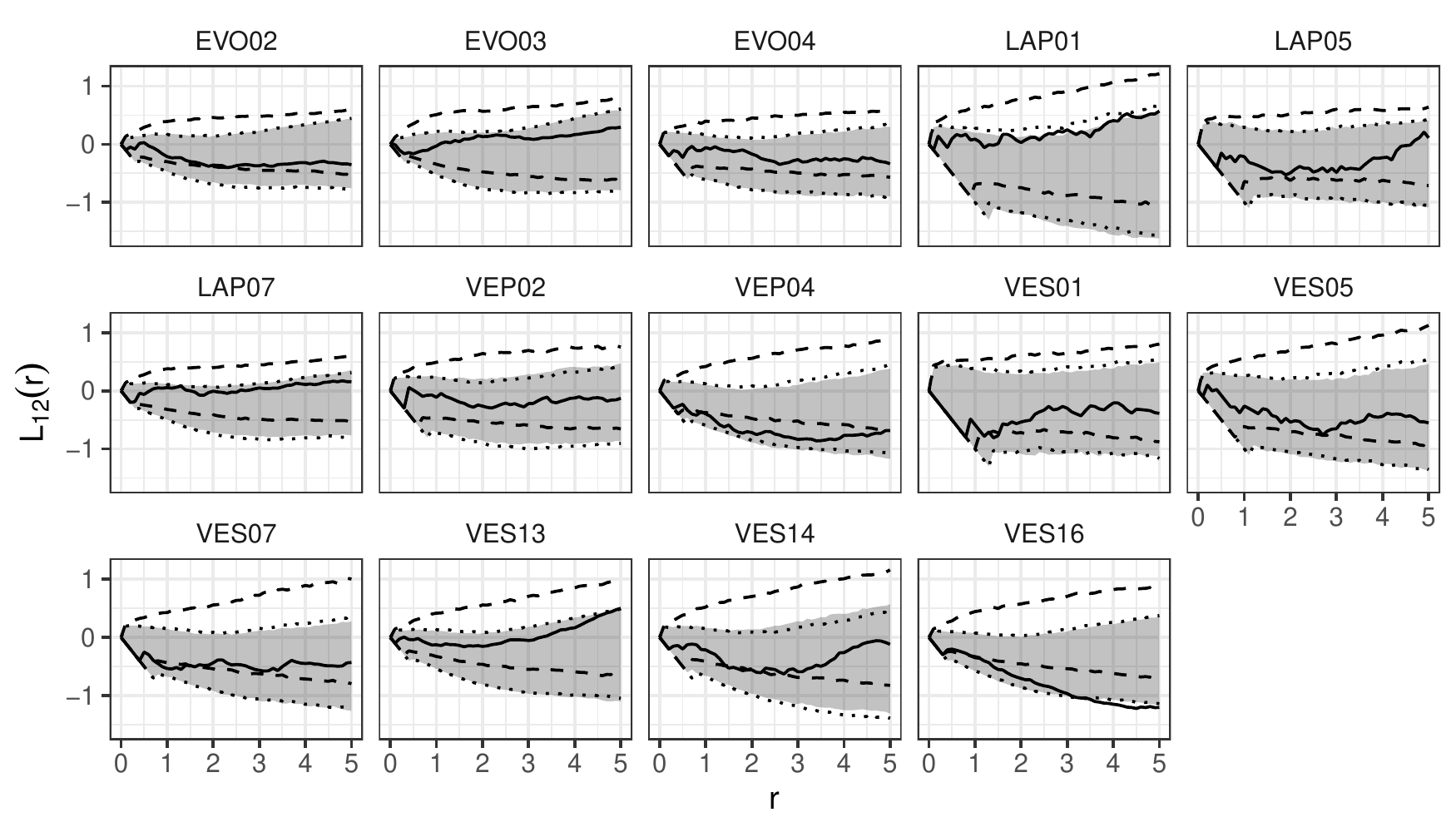}
\caption{Empirical $L_{12}$ functions (solid line) together with the 95\% global envelopes constructed from 10\,000 simulations from the posterior predictive distribution of the fitted conditional LGCP models for the 14 plots in Figure \ref{fig:seedlingppall} with mark independent \eqref{kernel_gaussian} (grey shade), mark dependent \eqref{kernel_Arne} (dotted lines), and no (dashed lines) influence.
}\label{fig:GETL12}
\end{figure}

Figure \ref{fig:GETL12} shows the empirical $L_{12}$ functions together with the 95\% global extreme rank length envelopes \citep{MyllymakiEtal2017, MyllymakiMrkvicka2020} constructed from the $L_{12}$ summary functions of the simulations of the fitted model with mark independent influence kernel \eqref{kernel_gaussian} (shaded region), mark dependent influence kernel \eqref{kernel_Arne} (dotted lines), and no influence kernel (dashed line) separately for each plot.
The observed $L_{12}$ function is distinctly better covered by the envelopes based on the models with influence field than without.
While the envelopes of the model without an influence field are centred around zero, i.e., no interaction between trees and seedlings, the empirical $L_{12}$ functions have the tendency to go below zero in most plots, indicating repulsion or inhibition of trees and seedlings, and the envelopes of the models with influence kernels are shifted downwards as well.
The difference between the two models with influence kernels is, however, minor.
Other summary functions ($L$, $F$, $G$) produced very similar envelopes regardless of the type or lack of influence field, see figures in Appendix \ref{appendix:Envelopes}.
The empirical functions were inside the envelopes, except the nearest neighbor distance distribution functions of four sample plots VES07, VES13, VES14 and VES16, which were slightly outside the envelopes at distances less than 1 m, i.e.\ less than the pixel size used in the discretization. This may suggest that the spatial distribution of the seedlings is not Poisson at a very small scale, but we did not investigate this further.

The envelopes for the models with mark dependent kernels with either $\delta$ or $\alpha$ set to zero are omitted because they were very similar to the envelopes of the other two influence kernels.

\begin{figure}[ht!]
\centering\includegraphics[width=\textwidth]{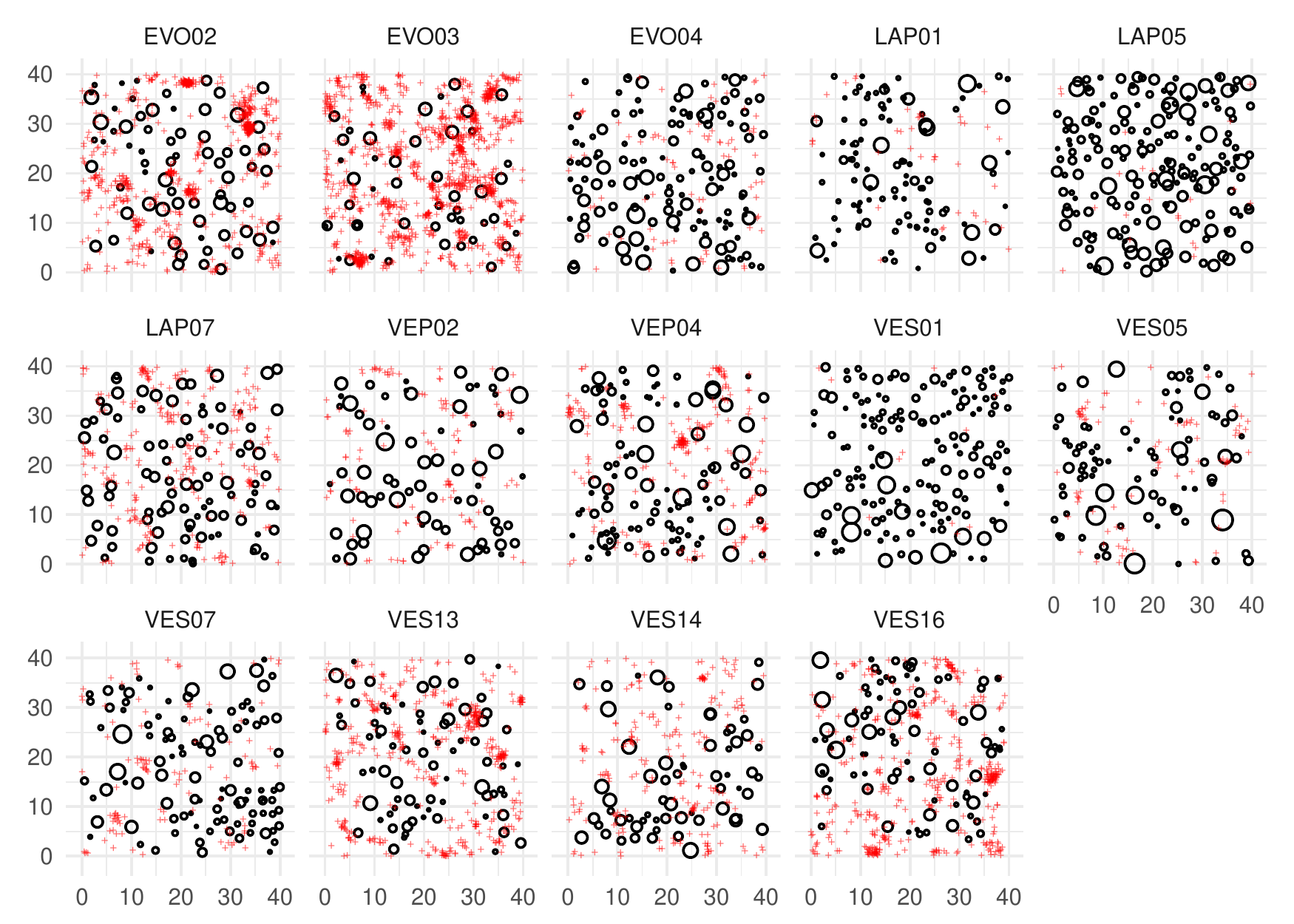}
\caption{Simulated seedling pattern (red crosses) and observed large trees (black circles, radius relative to dbh). The simulation was done using the posterior predictive distribution of the fitted conditional LGCP models for the 14 plots in Figure \ref{fig:seedlingppall} with mark independent influence of large trees.
}
\label{fig:simpp}
\end{figure}

\begin{figure}[ht!]
\centering\includegraphics[width=\textwidth]{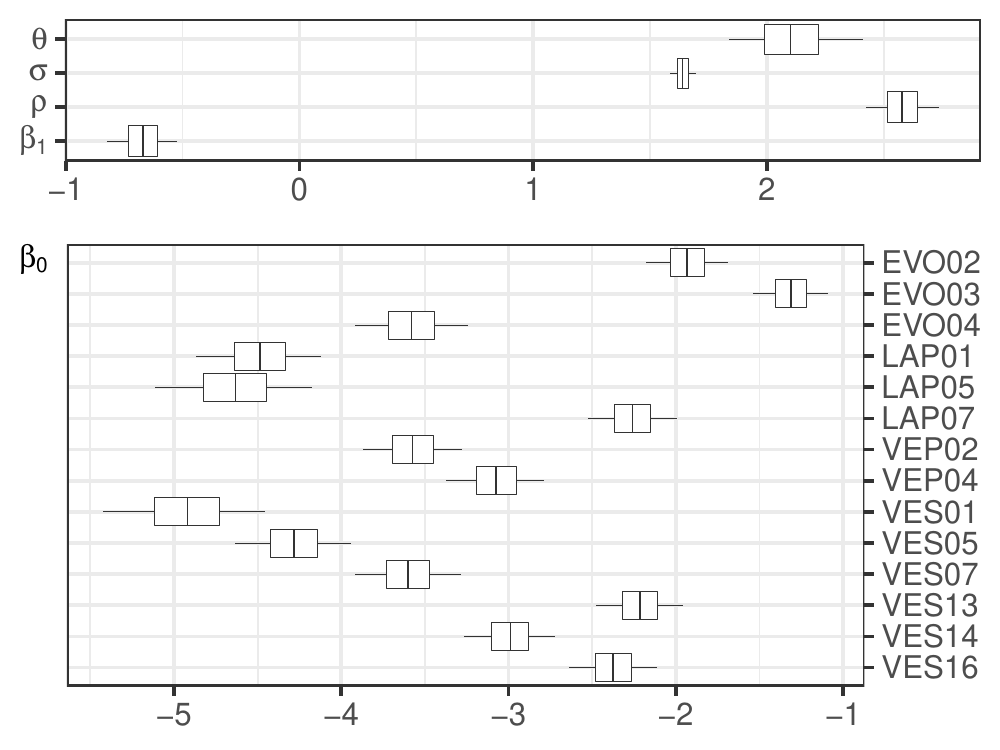}
\caption{Posterior quantiles (0.05, 0.25, 0.5, 0.75, 0.95) of the common parameters (top) and the sample plot specific intercepts $\beta_0$ (bottom).}\label{fig:posterior_marginals}
\end{figure}

Based on the analysis above, it is clear that an influence kernel is needed. However, since all the models with an influence kernel fitted the data equally well, we report the results of the simplest model \eqref{kernel_gaussian}.
The marginal posterior distributions of the model parameters of this model are shown in Figure \ref{fig:posterior_marginals}. The influence of the large trees on the seedlings ($\beta_1$) is clearly negative meaning that the seedlings avoid locations in the close vicinity of the large trees.
The range of influence $\theta$ of the large trees was estimated to be around 2.1 m, indicating that the influence of a large tree decreases from its maximum influence (at the tree location) to 37\% of it at distance 2.1 m from the tree, or to 5\% of it at distance 3.6 m.
However there is a lot of unexplained variability, as the quite wide envelopes in Figure \ref{fig:GETL12} and Appendix \ref{appendix:Envelopes} show.

Figure \ref{fig:simpp} shows for each plot one realisation drawn from the posterior predictive distribution of the model with mark independent influence kernel.
It is difficult to detect relationship between trees and seedlings by eye, but one can compare the clustering of the seedling patterns to the observed patterns (Figure \ref{fig:seedlingppall}). The patterns in Figure \ref{fig:seedlingppall} and Figure \ref{fig:simpp} look rather similar,
and according to the envelope tests (see Figure \ref{fig:GETL12} and Appendix \ref{appendix:Envelopes}) the model captures small scale structures up to 5 m distances.

\section{Discussion}\label{sec:discussion}

We proposed a
LGCP model to investigate the effect of large trees on the intensity of seedlings under the presence of unexplained clustering. The influence of large trees was modeled by using parametric influence kernels around them.
Our analysis suggests that tree regeneration is affected by the pattern of large trees in the studied data. Namely, the large trees were found to have negative effect on the seedling density in the vicinity of large trees.
Further, the LGCP model could capture much of the unexplained clustering.
For parameter estimation, we constructed a Bayesian approach using MCMC and Laplace approximation.
All computations were implemented in Julia language \citep{BezansonEtal2017}, while graphics were done using ggplot2 \citep{ggplot2}.

Estimation of the influence field parameters worked well in our simulation experiment and we did not observe any problems due to possible confounding between the influence field and the spatial random effect as reported in the literature \citep[e.g.][]{DupontEtal2020}. However,
the random field parameter estimates were biased. We suspect that this is caused by weak identifiability \citep[cf.][]{Anderes2010,Zhang2004} or discretization bias coupled with the Laplace approximation.
We used replicates to help with the weak identifiability which was necessary for the plots with  very few seedlings.
In our further experiments with finer discretizations (results not shown), we observed issues with the approximation. In particular, when the number of points per cell was small, the approximate posterior appeared to degenerate.
We are unaware of exact inference methods that would be feasible in our setting, but we are currently investigating new methods that could allow for more detailed investigation of this issue.

There are many alternative approaches to inference with log Gaussian Cox processes.
For example, the R package INLA \citep{RueEtal2009} uses Laplace approximation in a similar fashion as we did, but is somewhat restricted to linear models. Indeed, INLA can in principle accommodate our model using the \texttt{rgeneric} class (personal communication with Håvard Rue). However, we faced some computational difficulties in estimation. The R package lgcp \citep{TaylorEtal2015} uses MALA algorithm for efficient Bayesian inference for the full model including the latent field. The use of full MCMC might lead to better estimation of the random field parameters. However, the lgcp package is also restricted to linear models, whereby we were not able to apply it directly to our model. For Stan \citep{stan2018} our random field model appears to be too complicated, however there are some recent advances see e.g.\ \citet{MargossianEtal2020}. Also, inlabru \citep{inlabru}
could be further investigated.

Since the large trees outside the sampling window may affect the intensity of the seedlings within the window, an edge correction assuming that the large trees were from a Poisson process was included in the estimation procedure.
We demonstrated by a small simulation study that this edge correction can work well even when the large trees are from a regular process. Compared to no edge correction, it improved the parameter estimates
when the range of influence was rather wide.

Obvious alternative strategy would be to include the locations of the unobserved trees outside the observation window to the MCMC estimation in a similar manner as considered in the inference for Neyman-Scott point processes \citep{MollerWaagepetersen2004}.
This approach would allow incorporating prior information on the large tree process in the edge correction at the cost of increased complexity. Since we did not have important prior information and the effect of edge correction appeared minor, we did not explore this approach further.
Ideas from \citet{Geyer1999} or \citet{GabrielEtal2017} could be used to find further alternative edge correction methods.

Our proposed edge correction method can be efficiently implemented when the influence field is constructed as the sum of individual signals. In principle, a similar edge correction could be applied with different combination rules, such as product \citep[e.g.][]{WuEtal1985, MiinaPukkala2002} or max-fields \citep[e.g.][]{PenttinenNiemi2007}.
If also the influence kernel is binary, e.g.\ $c(h; \theta) = \mathbf{1}( h \leq \theta)$, then the max-field is split into two phases as well, namely influence and influence-free zones. However, we note that our proposed calculation of the expected influence of trees outside the observation window $W$ does not generalize directly to other combination rules.

The models introduced in this paper could be useful even for natural \citep[e.g.][]{AbellanasPerezMoreno2018} or urban forests \citep{HauruEtal2012}. Furthermore, they could be used in an experimental setup, where realizations of seedlings would be generated for different large tree patterns and the success of regeneration evaluated by some spatial summary functions such as the empty space function. In a similar manner, the effect of different thinning strategies on regeneration of trees could be evaluated.

It could be argued that, since the management was the same for all plots and the geographical differences minor, the plots should have had a common intercept parameter. However this was clearly not the case due to large variation in numbers of seedlings from plot to plot. Since we used plot specific intercepts, it could be argued that all other parameters should be plot specific too. This was not possible in practice due to the problems with the random field parameters. We did not explore the alternative where the intercept and the influence field parameters would be plot specific but the random field parameters shared since the envelope tests already suggested adequate fit of the model.

The observed and simulated seedling patterns in the Figures \ref{fig:seedlingppall} and \ref{fig:simpp}, respectively, are very similar in several aspects while quite different in others.
For example, the clusters seemed to be clustered in the VES13 plot.
Although the envelope tests suggest that the model was able to capture the variability in the data, it depends on the specific application if the model is adequate.
To best of our knowledge, this is the first point process model accounting for clustering of the seedlings in these uneven-aged forests.

There are many other factors than the vicinity of large trees that may affect the intensity of seedlings \citep{ValkonenMaguire2005,KuusinenEtal2019}. Therefore, the model could be further improved and unexplained variability decreased, if some covariate information on local conditions within plots would be available to be included in the model.
Further, plot level covariate effects could be added to the model in order to explain the numbers of seedlings in different plots.
Finally, we modeled the influence of large trees as a function of the dbh, whose effect on the influence was, however, minor in our data. Other possibly useful marks could be the height, crown ratio or crown width of the tree, for example.

\section*{Acknowledgements}

MK, MM and MV were financially supported by the Academy of Finland (Project Numbers 306875, 327211, 295100 and 315619) and AS by the Swedish Research Council (VR 2018-03986).
The authors thank Helena Henttonen, Jari Hynynen and Sauli Valkonen (Luke) for discussions on the application, and Antti Penttinen for his comments on an earlier version of the manuscript.
Further, the authors are grateful to Hilkka Ollikainen and Juhani Korhonen, who mastered the maintenance and measurements on the plots of the ERIKA data set.
The authors wish to acknowledge CSC – IT Center for Science, Finland, for computational resources.

\bibliographystyle{chicago}
\bibliography{Maris_bibfile}

\appendix
\section{Envelopes}\label{appendix:Envelopes}

\begin{figure}[ht!]
\centering\includegraphics[width=\textwidth]{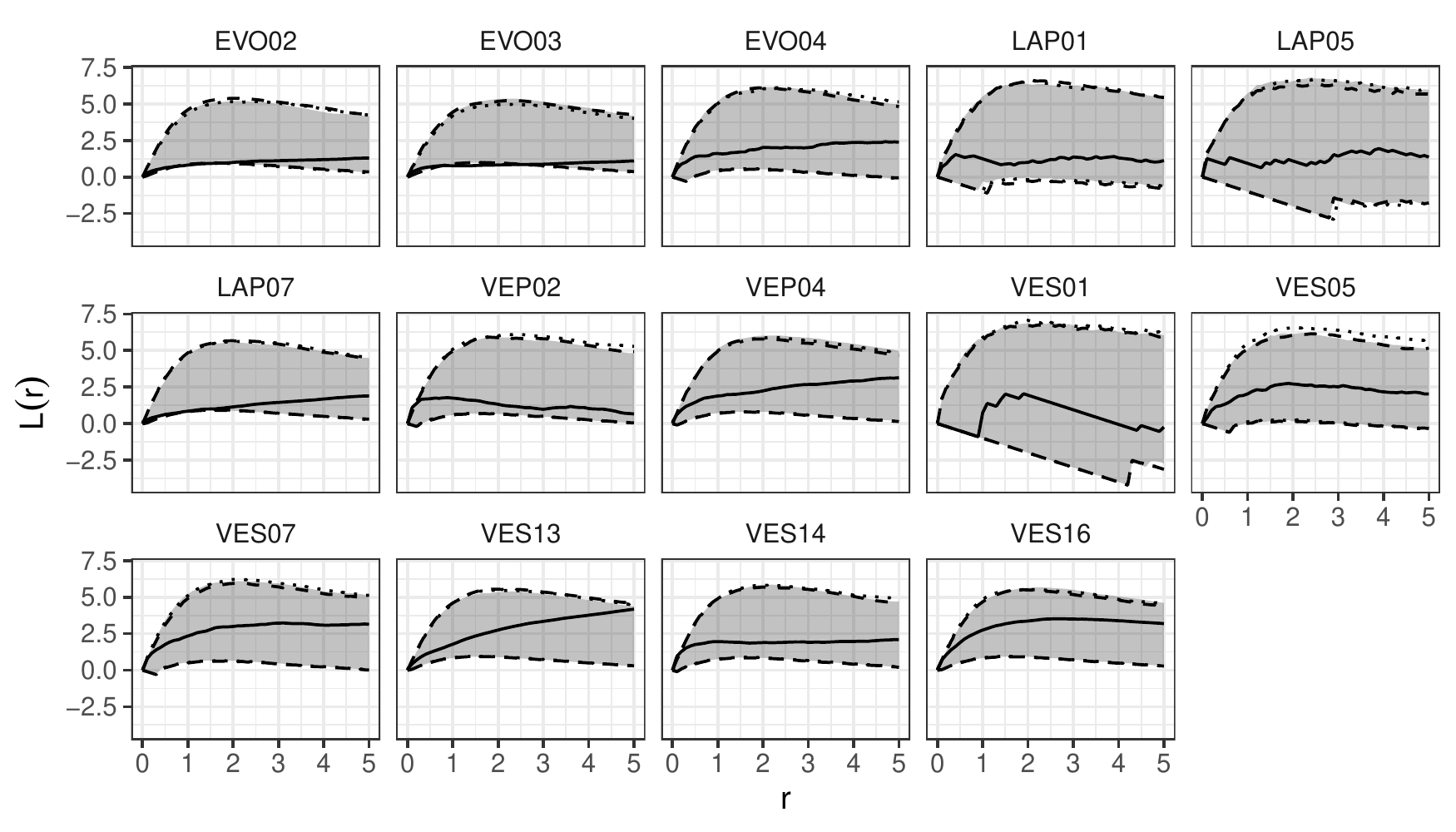}
\caption{Empirical $L$ functions (solid line) together with the 95\% global envelopes constructed from 10\,000 simulations from the posterior predictive distribution of the fitted conditional LGCP models for the 14 plots in Figure \ref{fig:seedlingppall} with mark independent (grey shade), mark dependent (dashed lines), and no (dotted lines) influence.
}
\end{figure}
\begin{figure}[ht!]
\centering\includegraphics[width=\textwidth]{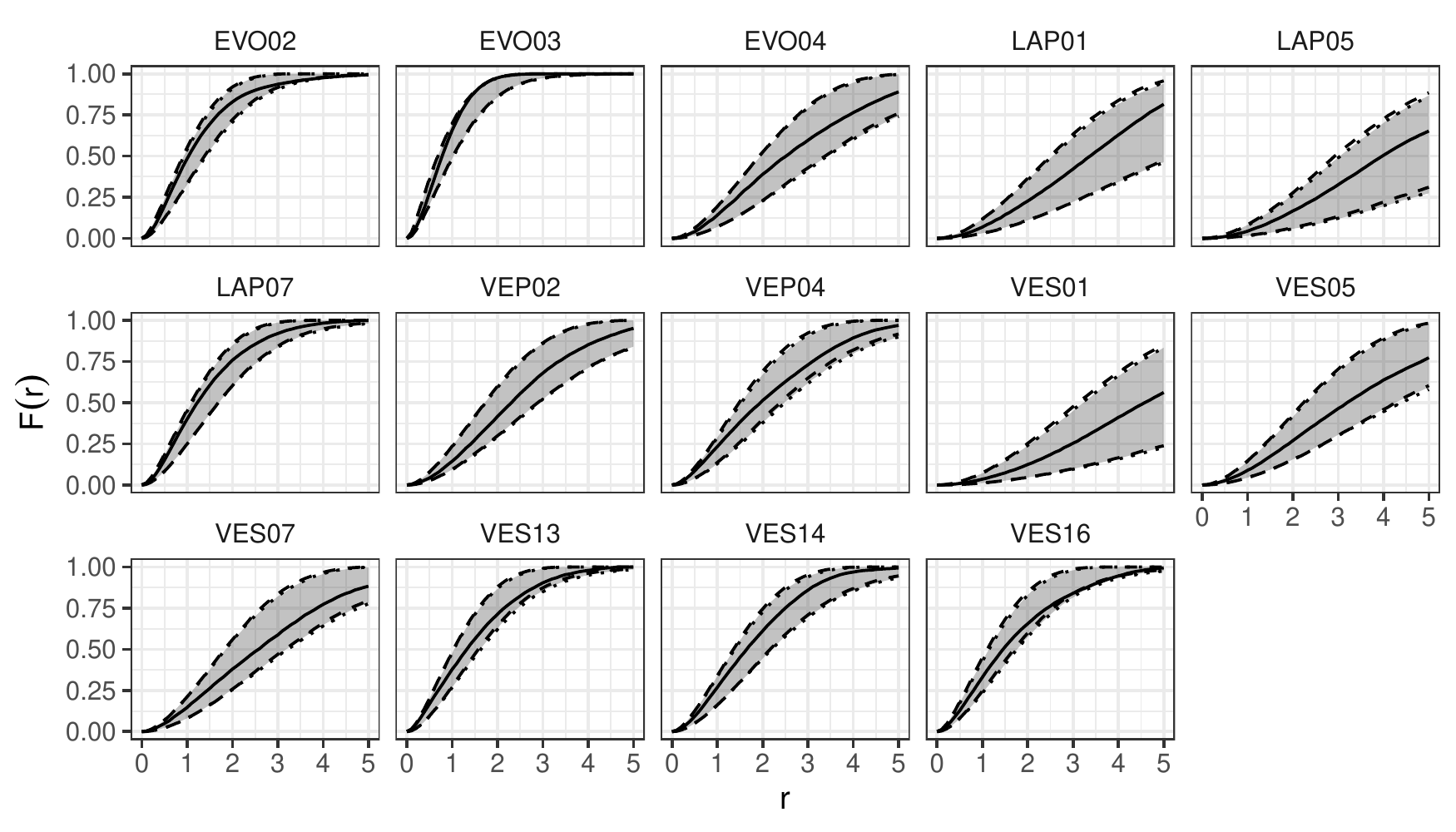}
\caption{Empirical empty space functions (solid line) together with the 95\% global envelopes constructed from 10\,000 simulations from the posterior predictive distribution of the fitted conditional LGCP models for the 14 plots in Figure \ref{fig:seedlingppall} with mark independent (grey shade), mark dependent (dashed lines), and no (dotted lines) influence.
}
\end{figure}
\begin{figure}[ht!]
\centering\includegraphics[width=\textwidth]{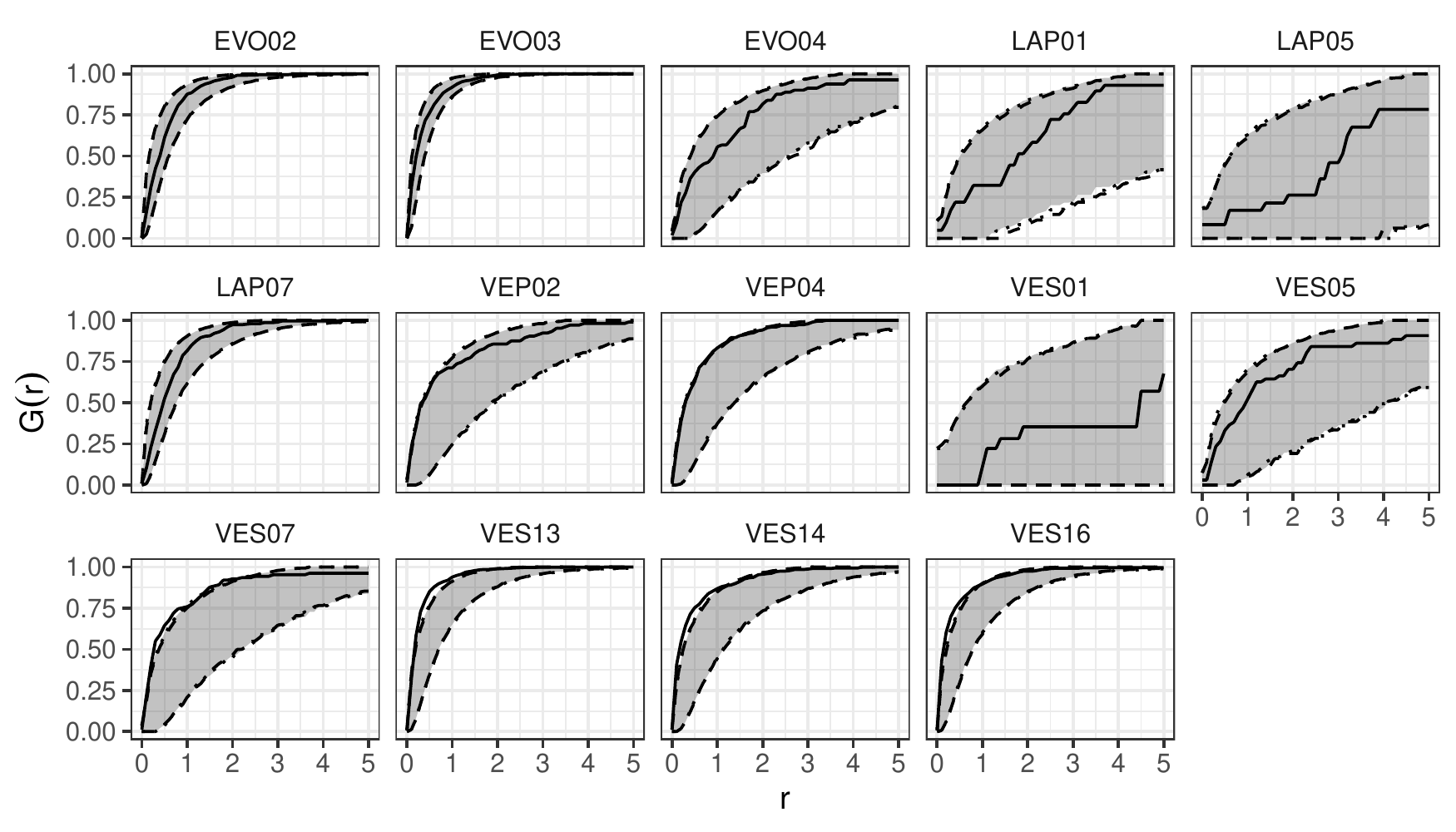}
\caption{Empirical nearest neighbor distance distribution functions (solid line) together with the 95\% global envelopes constructed from 10\,000 simulations from the posterior predictive distribution of the fitted conditional LGCP models for the 14 plots in Figure \ref{fig:seedlingppall} with mark independent (grey shade), mark dependent (dashed lines), and no (dotted lines) influence.
}
\end{figure}

\end{document}